\documentclass[a4paper,12pt]{article}
\usepackage{a4wide}
\usepackage[utf8]{inputenc}
\usepackage{graphicx}
\usepackage{amsmath}
\usepackage{amssymb}
\usepackage{array}
\usepackage{hyperref}
\usepackage{tabularx}
\usepackage{longtable}

\newcommand{\trace}[1]{\left\langle #1 \right\rangle}

\begin{document}

\newcolumntype{L}[1]{>{\raggedright\arraybackslash}p{#1}}
\newcolumntype{C}[1]{>{\centering\arraybackslash}p{#1}}
\newcolumntype{R}[1]{>{\raggedleft\arraybackslash}p{#1}}

 %%%%%%%%%%%%%%%%%%%%%%%%%%%%%%%%%%%%%%%%%%%%%%%%%%%%%%%%%%%%%%%%%%%%%%%%%%%
\begin{titlepage}
\setcounter{page}{0}
% flushright puts it towards the right of the page
\begin{flushright}
% our preprint number yy-nn (year-number)
LU TP 18-34\\
arXiv:1810.06834 [hep-ph]\\
% some indication of the date
October 2018 \\
\end{flushright}
\vfill
\begin{center}
%\title{ }

%\author{}
%\date{}
% put in line breaks to make the title look nicer and provide more
% space between the lines
{\large\bf The order $p^8$ mesonic chiral Lagrangian}
\\[3cm]
{\bf Johan Bijnens, Nils Hermansson-Truedsson and Si Wang}
\\[5mm]
{Department of Astronomy and Theoretical Physics, \\Lund University, S\"{o}lvegatan 14A, SE 223-62 Lund, Sweden }
\vfill

{\large\textbf{Abstract}}
\end{center}We derive the chiral Lagrangian at next-to-next-to-next-to-leading order (NNNLO)
for a general number $N_f$ of light quark flavours as well as for $N_f=2,3$.
We enumerate the contact terms separately. We also discuss the cases where some of the external fields are not included.
An example of a choice of Lagrangian is given in the supplementary material.

\end{titlepage}

%%%%%%%%%%%%%%%%%%%%%%%%%%%%%%%%%%%%%%%%%%%%%%%%%%%%%%%%%%%%%%%%%%%%%%%
\tableofcontents
\section{Introduction}
\label{sec:intro}

Chiral perturbation theory (ChPT)~\cite{Weinberg:1978kz,Gasser:1983yg,Gasser:1984gg} is an effective field theory (EFT) of QCD at low energies. An introduction and further references can be found in~\cite{Pich:2018ltt}. For $N_{f}$ flavours of light quarks the effective degrees of freedom are the  $N_{f}^2-1$ lightest pseudoscalar mesons. These arise as the (pseudo-)Goldstone bosons of the spontaneously broken chiral symmetry $SU(N_{f})_{L}\times SU(N_{f})_{R}$ to $SU(N_{f})_{V}$. For $N_{f}=2$, i.e. when only $u$ and $d$ quarks are considered, these are the pions, and for $N_{f}=3$ when the $s$ quark is included, the additional degrees of freedom are the kaons and the eta meson. The power counting is done in terms of a generic momentum, $p$, and, as in any EFT, a calculation is done to a fixed order in this variable. This requires a chiral Lagrangian of the corresponding order.
The leading order (LO), or $p^2$, next-to-leading order (NLO), or $p^4$, and next-to-next-to-leading order (NNLO), or $p^6$, chiral Lagrangians are known.
The LO is basically current algebra, the NLO was derived in~\cite{Gasser:1983yg,Gasser:1984gg}. The $p^4$ Lagrangian in the anomalous sector contains no new free parameters~\cite{Wess:1971yu,Witten:1983tw}.
The NNLO or $p^6$ Lagrangian was studied in~\cite{PhysRevD.53.315,Bijnens:1999sh}
and for the anomalous sector in~\cite{Bijnens:2001bb,Ebertshauser:2001nj}. 
In addition the divergence structure is known for all of these~\cite{Gasser:1983yg,Gasser:1984gg,Bijnens:2001bb,Bijnens:1999hw}.

The basis of~\cite{Bijnens:1999sh} for a general number of flavours and
$N_f=3$ is generally accepted as minimal. For two flavours, an extra
relation was found in~\cite{Haefeli:2007ty} and another one for the case of vanishing scalar and pseudoscalar external fields~\cite{Colangelo:2012ipa}. In general, it is quite difficult to be sure one has a minimal Lagrangian. The only foolproof method we are aware of is to find Green functions of the external fields that allow to determine all free parameters. This program has been done to NNLO for the case with vanishing scalar and pseudoscalar external fields in~\cite{Ruiz-Femenia:2015mia}. A very systematic approach to NNLO was done in the Diplom thesis~\cite{Weber:2008}. This latter reference was the inspiration for the present work.

In this paper we present a derivation of the next-to-next-to-next-to-leading order (NNNLO), or $p^8$, chiral Lagrangian $\mathcal{L}_{8}$ in the nonanomalous or positive intrinsic parity sector. We do this for two, three
and a general number of light quark flavours. We treat the cases of only Goldstone bosons, including scalar and pseudoscalar external fields,
including vector and axial vector external fields and including all four sets
of external fields.  As a byproduct we clarify the number of terms in the two-flavour case at NNLO.

There are several underlying reasons to do this work.
It is of general interest to know how many terms there actually are at a given order in the EFT expansions. For the case of extensions of the standard model there is a large literature, we quote here only the Hilbert series approach~\cite{Henning:2017fpj} and a software to find the minimal Standard Model EFT Lagrangian~\cite{Gripaios:2018zrz}. The remaining literature can be traced via these two. Our approach can in principle be extended to these cases as well. The second reason is that at tree level there are a number of works studying effective fields theories using amplitude methods~\cite{Cheung:2016drk} and references therein. Knowing the solution to $p^8$ provides a cross-check for extensions of that work to higher derivatives.
The third motivation to find the Lagrangian is that in a previous calculation~\cite{Bijnens:2017wba}, analytical expressions for the pion mass and decay constant at order $p^{8}$ were obtained for two flavour ChPT. However, as the NNNLO Lagrangian was not yet known, the tree level contributions were parametrised in terms of two unknown parameters.
These parameters are in fact combinations of low energy constants (LECs) multiplying the monomials in the minimal set we derive here, and their numerical values can in principle be found from experiment. The present status of their
values was reviewed in \cite{Bijnens:2014lea} and their determination from lattice QCD in \cite{FLAG}.

The Lagrangians are of the generic form
\begin{align}
\mathcal{L}_{2n} = \sum_i c_i^{(2n)} \mathcal{O}_i^{(2n)}\,.
\end{align}
The $c_i^{(2n)}$ are the free constants, usually referred to as low-energy-constants (LECs). The $\mathcal{O}_i$ are monomials in terms of the (pseudo-)Goldstone bosons and
external fields that are invariant under chiral symmetry, hermitian and invariant under charge conjugation and parity\footnote{Since we have a field theory this also implies invariance under time-reversal because of the $CPT$-theorem.}. The set of $\mathcal{O}_i$ should be complete and minimal.

We first write down all possible single term operators\footnote{For these we use the term operators in the remainder.} in terms of a number of basic building blocks. At this level we impose parity ($P$) and invariance under chiral symmetry. This is described in section~\ref{sec:chirallag1}.
This basis is the simplest one to derive all the relations described below.
At this stage we also take into account the different ways to write each operator due to cyclicity of the trace, relabelling of indices and those that are zero because of contraction of antisymmetric and symmetric pairs. The next step is then to combine these operators into monomials that are invariant under charge conjugation ($C$) and hermitian conjugation ($H$).
The method we use here is discussed in section~\ref{sec:chirallag2}. An extension was needed to find all possible contact terms, i.e. the operators only depending on external fields. This is discussed in section~\ref{sec:chirallag3}.

In general this leads to a very large number of operators, but here the importance is to be complete. It is also easier to make sure we have obtained all relations when the basis at this level contains all possible single term operators, even those that can obviously be removed. The linear relations between operators that we use are all partial integrations, finding operators that can be removed using field redefinitions by using the equation of motion,
the Bianchi identity
and relations that follow from commuting derivatives. These are discussed in
sections~\ref{sec:relations1}-\ref{sec:relations4} and are what we use to determine the general $N_f$ flavour Lagrangian. For two- and three flavours there are additional relations following from the Cayley-Hamilton relation. These are discussed in section~\ref{sec:relations5}. Again, there are a number of extra conditions needed when determining the relations between possible contact terms. These are discussed in section~\ref{sec:relations6}.

The analytical work is done using \textsc{FORM}~\cite{Vermaseren:2000nd} and \textsc{Python} to rewrite some of \textsc{FORM}'s output back into \textsc{FORM} commands. Identical relations are removed within the \textsc{FORM} programs but we still end up with a very large number of operators and relations. The number of independent relations was determined by using Gaussian elimination. This we implemented using \textsc{C++} with
exact arithmetic via the \textsc{GNU} multiple precision library (\textsc{GMP})~\cite{Granlund:2016}.

The results are discussed in section~\ref{sec:nnnlolag}. We give here the numbers of independent monomials for the various cases of number of flavours and inclusion of external fields. Explicit lists of the monomials involved are relegated
to the appendix for the contact terms and to the supplementary material for the others. As an example of the use of our Lagrangians we discuss meson-meson scattering in section~\ref{sec:pipi} and show that the number of LECs agrees with general considerations on the form of the amplitude.

\section{Building blocks for chiral Lagrangians}
\label{sec:chirallag}
ChPT is constructed from the chiral symmetry $G= SU(N_{f})_{L}\times SU(N_{f})_{R}$ of the QCD Lagrangian with $N_{f}$ massless quarks. Due to a non-vanishing quark vacuum expectation value, this symmetry is spontaneously broken to $H=SU(N_{f})_{V}$ and the corresponding $N_{f}^{2}-1$ Goldstone bosons live in the coset space $G/H$ and are the lightest pseudoscalar mesons.

Gasser and Leutwyler~\cite{Gasser:1983yg,Gasser:1984gg} added scalar $s$, pseudoscalar $p$, vector $v_\mu$ and
axial-vector $a_\mu$ external fields to the QCD Lagrangian. They transform in such a way that the chiral group $G$ can be made local and leave the QCD Lagrangian
\begin{align}
\mathcal{L} = \mathcal{L}^0_{QCD}+
\overline q\gamma^\mu\left(v_\mu+a_\mu\gamma_5\right)-\overline q \left(s-ip\gamma_5\right)q
\end{align}
invariant. $q$ is an $N_f$ column vector, $s,p,v_\mu,a_\mu$ are $N_f\times N_f$ matrices in flavour space. We assume that $v_\mu$ and $a_\mu$ are traceless.

The Goldstone boson fields $\phi$ can be organized in a unitary matrix $u(\phi)$~\cite{Coleman:1969sm,Callan:1969sn}. The transformation under
a chiral symmetry transformation $(g_L,g_R)\in SU(N_f)_L\times SU(N_f)_R$
of the relevant fields is
\begin{align}
\label{eq:transformation}
u(\phi)&\,\longrightarrow g_R u(\phi) h(g_L,g_R.u(\phi))^\dagger = h(g_L,g_R.u(\phi))u(\phi)g_L^\dagger\,,
\nonumber\\
\chi\equiv  2B\left(s+ip\right)&\,\longrightarrow g_R\chi g_L^\dagger\,,
 \nonumber\\
\ell_\mu\equiv v_\mu-a_\mu&\,\longrightarrow g_L \ell_\mu g_L^\dagger-i\partial_\mu g_L g_L^\dagger\,,
\nonumber\\
r_\mu\equiv v_\mu+a_\mu&\,\longrightarrow g_R \ell_\mu g_R^\dagger-i\partial_\mu g_R g_R^\dagger\,.
\end{align}
The top line in (\ref{eq:transformation}) is the definition of $h(g_L,g_R,u(\phi))$ which is an element of the unbroken subgroup $H$.
The remaining lines in (\ref{eq:transformation}) define the combinations $\chi,\ell_\mu$ and $r_\mu$.
In the remainder we drop the arguments of $u$ and $h$.
The constant $B$ in the definition of $\chi$ is related to the chiral limit
value of the quark vacuum expectation value and pion decay constant
$B=-\langle \bar{q}q\rangle /N_{f}F^{2}$. We also use the notation
$\langle X\rangle = tr\left(X\right)$ for any matrix in flavour space.
The effective Lagrangian is expanded in powers of $p^{2}$ according to
\begin{align}
\mathcal{L} = \mathcal{L}_{0}+  \mathcal{L}_{4} +   \mathcal{L}_{6}+ \mathcal{L}_{8}+\ldots \, ,
\end{align}
where the first four contributions are the LO, NLO, NNLO and NNNLO Lagrangians, respectively. Each of these is a sum of order $p^{2n}$ monomials $\mathcal{O}_{i}^{(2n)}$, i.e.
\begin{align}
\mathcal{L}_{2n} = \sum _{i=1}^{N_{2n}}c_{i}^{(2n)}\mathcal{O}_{i}^{(2n)}\, ,
\end{align}
where the $c_{i}^{(2n)}$ are LECs and $N_{2n}$ is the number of monomials.
We use here the standard counting where the fields $s,p$ are counted as $p^2$,
$v_\mu,a_\mu$ as order $p$.
The monomials $\mathcal{O}_{i}^{(2n)}$ depend on the field content, and each is invariant under chiral symmetry, Lorentz transformations and the discrete transformations parity, charge conjugation and hermitian conjugation.

The easiest way to construct invariants is to produce objects that transform linearly under a simple symmetry group. Here one can use different choices,
all transforming purely left or purely right-handed or purely under the vector
group $SU(3)_V$. The former choices were made in \cite{PhysRevD.53.315,Ebertshauser:2001nj,Bijnens:1989jb}, the latter in
\cite{Bijnens:1999sh,Bijnens:2001bb}. We will use the latter except for the determination of the contact terms.

\subsection{Chiral building blocks}
\label{sec:chirallag1}

The $p^4$ Lagrangian was constructed in~\cite{Gasser:1984gg} using
\begin{align}
%\label{eq:defU}
%\label{eq:fr}
%\label{eq:fl}
U\equiv u^2&\longrightarrow  g_{R}  U  g_{L}^\dagger \,,
\nonumber\\
\chi&\longrightarrow  g_{R}  \chi  g_{L}^\dagger \,,
\nonumber\\
 F_{L}^{\mu \nu} \equiv \partial ^{\mu}\ell ^{\nu}-\partial ^{\nu}\ell ^{\mu} -i\left[ \ell ^{\mu},\ell ^{\nu} \right]&\,\longrightarrow
g_L  F_{L}^{\mu \nu} g_L^\dagger\,,
\nonumber\\
 F_{R}^{\mu \nu} \equiv \partial ^{\mu}r ^{\nu}-\partial ^{\nu}r ^{\mu} -i\left[ r ^{\mu},r ^{\nu} \right] &\,\longrightarrow
g_R  F_{R}^{\mu \nu} g_R^\dagger\,.
\end{align}
We will refer to $F_{L,R}^{\mu \nu}$ as field strengths.

We can define covariant derivatives $D_\mu$ acting on these building blocks that transform as the building blocks themselves via 
\begin{align}\label{eq:covder}
D_{\mu} O =
\left\{
\begin{array}{ll}
 \partial _{\mu}O -ir_{\mu}O +iO \ell _{\mu},  & O\longrightarrow g_{R}\, O\, g_{L}^{\dagger}  
\, ,  \\ \\ 
\partial _{\mu}O  -i \ell _{\mu}O+iOr_{\mu},  & O\longrightarrow g_{L}\, O\, g_{R}^{\dagger}  
\, ,  \\ \\
\partial _{\mu}O -ir_{\mu}O +i O r _{\mu},  & O\longrightarrow g_{R}\, O\, g_{R}^{\dagger}  
\, ,  \\ \\
\partial _{\mu}O -i \ell_{\mu}O +i O\ell _{\mu},  & O\longrightarrow g_{L}\, O\, g_{L}^{\dagger}  
\, ,  \\
\end{array}
\right. 
\end{align}
for any operator $O$ transforming in one of the four ways. The respective momentum orders of the above structures are $U\sim p^{0}$, $D_{\mu}U \sim p$ and $\chi, F_{L,R}^{\mu \nu}\sim p^{2}$. Their transformation properties under the discrete transformations can be found in table~\ref{table:transpropsLR}.
This set of building blocks we will refer to as the first basis.
\begin{table}%[tbh]
  \begin{center}  
  \renewcommand{\arraystretch}{1.2}
    \begin{tabular}{|C{2cm}|C{3cm}|C{3cm}|C{3cm}|}
      \hline
                        & $P$ & $C$ & h.c. \\ \hline 
      $U$         & $U ^{\dagger}$ &  $U ^{T}$& $U ^{\dagger}$ \\
      $\chi$         & $\chi ^{\dagger}$ &  $\chi ^{T}$& $\chi ^{\dagger}$ \\
      $F_{L}^{\mu \nu }$     & $\varepsilon(\mu)\varepsilon(\nu)F_{R}^{\mu \nu}$ & $-\big( F_{R}^{\mu \nu}\big) ^{T}$ & $F_{L}^{\mu \nu}$ \\
      $F_{R}^{\mu \nu}$     & $\varepsilon(\mu)\varepsilon(\nu)F_{L}^{\mu \nu }$ & $-\big( F_{L}^{\mu \nu}\big) ^{T}$& $F_{R}^{\mu \nu}$ \\
     \hline
    \end{tabular}
   \renewcommand{\arraystretch}{1}
     
    \caption{Transformation properties of the first chiral building blocks. $\varepsilon(0)=-\varepsilon(i=1,2,3)=1$.}\label{table:transpropsLR}
  \end{center}
\end{table}

As mentioned above it is easier to use building blocks that all transform the same way. We choose to use the transformation under $h\in H$ as defined in (\ref{eq:transformation}) and all building blocks $X$ transforming as
$X\longrightarrow h X h^\dagger$.
It can be checked that the following blocks transform in the desired way and form a complete set:
\begin{align}
\label{eq:defblocks}
 u_{\mu}&\, = i\Big[ u^{\dagger} (\partial _{\mu}-ir_{\mu})u-u(\partial _{\mu}-i\ell _{\mu} )u^{\dagger}\Big]
\, ,\nonumber \\
%\label{eq:chipm}
 \chi _{\pm}&\, = u^{\dagger}\chi u^{\dagger}\pm u\chi ^{\dagger} u
\, , \nonumber\\
%\label{eq:fpm}
 f_{\pm}^{\mu \nu}&\, = u F_{L}^{\mu \nu }u^{\dagger} \pm u^{\dagger}F_{R}^{\mu \nu}u
 \, .
% \\
%\label{eq:h}
%& h_{\mu \nu} = \nabla _{\mu} u_{\nu}+\nabla _{\nu} u_{\mu} 
%\, , \\
%\label{eq:chipm}
%& \chi _{\pm \, \mu} = u^{\dagger} D_{\mu} \chi \, u^{\dagger}\pm u D_{\mu}\chi ^{\dagger}\, u = \nabla _{\mu} \chi _{\pm}-\frac{i}{2}\left\{ \chi _{\mp},u_{\mu}\right\}
%\, .
\end{align}
Note that when no external fields are included, only $u_{\mu}$ with $\ell _{\mu}=r_{\mu} = 0$ needs to be considered. The covariant derivative $\nabla _{\mu}  $ is defined as
\begin{align}\label{eq:nablader}
& \nabla _{\mu} X= \partial _{\mu}X +\left[ \Gamma _{\mu}, X\right] 
\, , 
\end{align}
where the chiral connection is
\begin{align}
\Gamma _{\mu} = \frac{1}{2}\Big[ u^{\dagger}\left( \partial _{\mu}-ir_{\mu}\right) u+u\left( \partial _{\mu}-i\ell _{\mu}\right) u^{\dagger} \Big] \, .
\end{align}
Equations~(\ref{eq:covder}) and~(\ref{eq:nablader}) highlight one of the reasons this set of building blocks is more convenient than the first set. We will refer to this as the main set in the remainder.

Since $\nabla _{\mu}X$ transforms like $X$ under chiral symmetry, we can always take covariant derivatives of the structures in~(\ref{eq:defblocks}) and obtain another building block of one order higher in $p$-counting. The respective orders of the above structures are $u_{\mu},\nabla_\mu\sim p$ and $%h_{\mu \nu}, 
\chi _{\pm}, f_{\pm \mu \nu} \sim p^{2}$. Their transformation properties under $P$, $C$ and hermitian conjugation are given in table~\ref{table:transprops}. Further note that $u_{\mu}$ %, $h_{\mu \nu}$
and $f_{\pm \mu \nu}$ are traceless since $r_{\mu}$ and $\ell _{\mu}$ are. 
\begin{table}
  \begin{center}  
  \renewcommand{\arraystretch}{1.5}
    \begin{tabular}{|C{2cm}|C{3cm}|C{3cm}|C{3cm}|}
      \hline
                        & $P$ & $C$ & h.c. \\ \hline 
      $u_{\mu}$         & $-\varepsilon (\mu) u_{\mu}$ &  $u_{\mu}^{T}$& $u_{\mu}$ \\
%      $h_{\mu \nu}$     & $-\varepsilon (\mu)\varepsilon (\nu) h_{\mu \nu}$ & $h_{\mu \nu}^{T}$ & $h_{\mu \nu}$ \\
      $\chi _{\pm}$     & $\pm \chi _{\pm}$ & $\chi _{\pm}^{T}$& $\pm \chi _{\pm}$ \\
      $f_{\pm \mu \nu}$ &  $\pm\varepsilon (\mu)\varepsilon (\nu) f_{\pm \mu \nu}$ & $\mp f_{\pm \mu \nu}^{T}$ & $f_{\pm \mu \nu}$ \\ \hline
    \end{tabular}
   \renewcommand{\arraystretch}{1}
     
    \caption{Transformation properties of the main chiral building blocks, where $\varepsilon (0) = -\varepsilon (i=1,2,3) = 1$.}\label{table:transprops}
  \end{center}
\end{table}

All possible operators of a given order in $p$ that are invariant under the chiral group can be generated by taking traces of the building blocks and their covariant derivatives. At this stage we make sure that terms that are the same from cyclicity of traces or relabelling indices are identified.
We also remove terms that are obviously zero because a pair of antisymmetric indices is contracted with a pair of symmetric ones.

We look here at the nonanomalous sector so we
do not need the Levi-Civita tensor $\varepsilon_{\mu\nu\alpha\beta}$.
Lorentz-indices are always contracted pairwise so parity can be easily
implemented as well by keeping only terms with an even number of $u_\mu$, $\chi_-$
and $f_{-\mu\nu}$.

\subsection{A basis of odd and even monomials}
\label{sec:chirallag2}

In addition to chiral invariance and parity as implemented in the single term operators built directly from the building blocks we need to impose charge conjugation and have terms that are hermitian. One advantage of the building blocks is that charge conjugation and hermitian conjugation always relate the same operators. This can be seen from
\begin{align}
C\left(\langle X_1\ldots X_n\rangle\right) 
&\,= \pm\langle X_1^T\ldots X_n^T\rangle =\pm \langle X_N\ldots X_1\rangle\,,
\nonumber\\
\left(\langle X_1\ldots X_n\rangle\right)^\dagger 
&\,= \langle X_n^\dagger\ldots X_1^\dagger\rangle =\pm \langle X_N\ldots X_1\rangle\,,\end{align}
where $X_i$ is any of the building blocks in (\ref{eq:defblocks}) or their
covariant derivatives.

We can now construct the combinations that are hermitian and odd or even under $C$. Any operator transforms according to
\begin{align}
\mathcal{O}_{i}\longrightarrow \lambda ^{{C}}_{\pm}\, \lambda ^{\textrm{h.c.}}_{\pm} \, \mathcal{O}_{j} \, ,
\end{align} 
where $\mathcal{O}_{j}$ can be any operator in the basis and $\lambda _{\pm}^{{C},\textrm{h.c.}}=\pm 1$ are the eigenvalues under the respective discrete transformations. There are thus four possibilities of the signs, which we denote by $(\lambda ^{{C}}_{\pm},\lambda ^{\textrm{h.c.}}_{\pm})$,
in addition we can have $i=j$ or $i\ne j$.
We define ${C}$-even and ${C}$-odd monomials $\mathcal{O}^{\pm}_{ i}$ for the case $j=i$:
\begin{align}
 & (+,+): \; \; \;  \mathcal{O}_{i} = \mathcal{O}^{+}_{i} \, , \nonumber \\
 & (-,+): \; \; \;   \mathcal{O}_{i} = \mathcal{O}^{-}_{i} \, , \nonumber \\
 & (+,-): \; \; \;  \mathcal{O}_{i} = i\mathcal{O}^{+}_{i} \, , \nonumber \\
 & (-,-): \; \; \;  \mathcal{O}_{i} = i\mathcal{O}^{-}_{i} \, . 
\end{align}
For $j\neq i$ we instead have
\begin{align}
  (+,+): \; \; \; & \mathcal{O}_{i} = \frac{\mathcal{O}^{+}_{i} +i\mathcal{O}^{-}_{i}}{2}\, ,
%\nonumber
%\\  
\; \; \; 
%&
 \mathcal{O}_{j} =  \frac{\mathcal{O}^{+}_{i} -i\mathcal{O}^{-}_{i}}{2} \, , \nonumber 
\\
  (-,+): \; \; \; & \mathcal{O}_{i} = \frac{i\mathcal{O}^{+}_{i} +\mathcal{O}^{-}_{i}}{2}\, ,
%\nonumber
%\\  
\; \; \; 
%&
 \mathcal{O}_{j} =  \frac{-i\mathcal{O}^{+}_{i} +\mathcal{O}^{-}_{i}}{2} \, , \nonumber
\\  
  (+,-): \; \; \; & \mathcal{O}_{i} = \frac{i\mathcal{O}^{+}_{i} +\mathcal{O}^{-}_{i}}{2}\, ,
%\nonumber
%\\ 
\; \; \; 
%&
 \mathcal{O}_{j} =  \frac{i\mathcal{O}^{+}_{i} -\mathcal{O}^{-}_{i}}{2} \, , \nonumber 
\\
  (-,-): \; \; \; & \mathcal{O}_{i} = \frac{\mathcal{O}^{+}_{i} +i\mathcal{O}^{-}_{i}}{2}\, ,
%\nonumber
%\\
\; \; \;
% &
 \mathcal{O}_{j} =  \frac{-\mathcal{O}^{+}_{i} +i\mathcal{O}^{-}_{i}}{2} \, .
\end{align}
The final Lagrangian should only contain the monomials $\mathcal{O}^+_i$.

\subsection{Contact terms}
\label{sec:chirallag3}

The monomials constructed using the main building blocks in (\ref{eq:defblocks})
are sufficient to construct a complete basis and determine the total number of terms in the Lagrangian. However, there are in general terms possible that only depend on the external fields, i.e. the contact terms. These cannot be
simply measured in physical processes but depend on the precise definitions of the external fields used in QCD. For determining the number of physically relevant parameters at a given order it is therefore desirable to also know the number of contact terms.
 
The contact terms only depend on external fields and are therefore best handled in the first basis using $O=\chi, F_L^{\mu\nu},F_R^{\mu\nu}$ and their covariant derivatives. The relations between the building blocks in this and the main basis are
\begin{align}
\label{eq:contactblocks}
 \chi &\,= \frac{1}{2}u\left( \chi _{+}+\chi _{-}\right) u 
\, , \nonumber\\
 \chi^{\dagger}&\, = \frac{1}{2}u^{\dagger}\left( \chi _{+}-\chi _{-}\right) u^{\dagger} 
\, ,\nonumber \\
 F_L^{\mu\nu}&\, = \frac{1}{2}u^{\dagger} \left( f_{+}^{\mu \nu}+f_{-}^{\mu \nu}\right)u
 \, ,\nonumber \\ 
 F_R^{\mu \nu}&\, = \frac{1}{2}u \left( f_{+}^{\mu \nu}-f_{-}^{\mu \nu}\right)u^\dagger \, .
\end{align}
This shows that the contact terms are included when using the main basis.

The contact terms are found by writing down all possible chiral invariant trace structures of the building blocks $O$ and their covariant derivatives, taking into account that the transformations under the left and right group should match. One also needs to be more careful with defining the $P,C$ and hermitian conjugation even operators generalizing the method of the previous subsection.

The third complication is that there is for a given number of flavours another combination with a well-defined chiral transformation~\cite{Kaplan:1986ru}:
\begin{align}
\tilde\chi\equiv\left(\det(\chi)\chi^{-1}\right)^\dagger
\longrightarrow g_R \tilde\chi g_L^\dagger\,.
\end{align}
This allows to take care of the invariant operator
$\textrm{det} \, \chi + \textrm{det} \, \chi ^{\dagger}$ of chiral order $p^{2N_{f}}$ and the extra building block for the two-flavour case in~\cite{Gasser:1983yg}, called $\tilde\chi$ there.

$\tilde\chi$ is perfectly regular when $s,p$ go to zero. This can be seen from
the explicit expressions using $\chi_{ij} = x_{ij}$:
\begin{align}
\label{eq:chitilde}
&N_f=2: &
% \chi &\,=\begin{pmatrix}x_{11} & x_{12} \\x_{21} & x_{22}\end{pmatrix}&
% &\Longrightarrow&
 \tilde\chi&\,=\begin{pmatrix}x^*_{22} & -x_{21}^*\\-x^*_{12} & x_{11}^*\end{pmatrix}\,,
\nonumber\\
&N_f=3: &
% \chi &\,=\begin{pmatrix}x_{11} & x_{12} & x_{13}\\
%                         x_{21} & x_{22} & x_{23}\\
%                         x_{31} & x_{32} & x_{33}\end{pmatrix}&
% &\Longrightarrow&
 \tilde\chi&\,=\begin{pmatrix}x^*_{22}x^*_{33}-x^*_{23}x^*_{32} &
                              x^*_{31}x^*_{23}-x^*_{21}x^*_{33} &
                              x^*_{21}x^*_{32}-x^*_{31}x^*_{22} \\
                              x^*_{32}x^*_{13}-x^*_{12}x^*_{33} &
                              x^*_{11}x^*_{33}-x^*_{31}x^*_{13} &
                              x^*_{31}x^*_{12}-x^*_{11}x^*_{32} \\
                              x^*_{12}x^*_{23}-x^*_{22}x^*_{13} &
                              x^*_{21}x^*_{13}-x^*_{11}x^*_{23} &
                              x^*_{11}x^*_{22}-x^*_{12}x^*_{21}
                            \end{pmatrix}
\end{align}
Terms involving $\tilde\chi$ can be rewritten in the other building blocks so they are included
in our main basis. However, structures involving $\tilde\chi$ are not fully independent  of structures with $\chi$, so a careful check that they are really
independent is needed. An example is
 $\langle\tilde\chi\tilde\chi^\dagger\rangle=\langle\chi\chi^\dagger\rangle$ for $N_f=2$. The easiest way to check is simply to use the explicit forms given
in (\ref{eq:chitilde}).

\section{Constructing the chiral Lagrangian}
\label{sec:relations}

Equipped with the chiral building blocks of the main set basis it is now possible to construct the Lagrangians. By demanding that the operators be invariant under chiral symmetry, Lorentz transformations and the discrete symmetries of QCD, the only possible monomials at LO are
\begin{align}
\langle u_\mu u^\mu\rangle,\quad\langle\chi_+\rangle\,.
\end{align}
This gives the LO Lagrangian in the standard form
\begin{align}\label{eq:LOLag}
\mathcal{L}_{0} = \frac{F^{2}}{4}\langle u_{\mu}u^{\mu}+\chi _{+}\rangle \, .
\end{align}
$F$ is a LEC corresponding to the LO pion decay constant and the second LEC $B$ is hidden inside $\chi$ as given in (\ref{eq:transformation}).

At higher orders many more structures must be included, and it is important to note that the operators $\mathcal{O}_{i}$ can be linearly related. The relations relevant for the NNNLO Lagrangian will be discussed in more detail below, but in short they arise from
\begin{itemize}
\item the vanishing of total derivatives (i.e. from integration by parts),
\item terms that can be removed using field redefinitions,
this is equivalent to removing combinations of operators that vanish
due to the LO equation of motion (EoM), for a proof see~\cite{Scherer:1994wi} and appendix A of~\cite{Bijnens:1999sh},
\item the Bianchi identity of the field-strength tensor $\Gamma _{\mu \nu} = \frac{1}{4}\left[ u_{\mu}, u_{\nu}\right] -\frac{i}{2}f_{+\mu \nu} $,
\item the identities $f_{-\mu \nu}-\nabla _{\nu} u_{\mu}+\nabla _{\mu} u_{\nu} = 0 $ and $\left[ \nabla _{\mu},\nabla _{\nu} \right] X = \left[ \Gamma _{\mu \nu},X\right]  $,
\item the Cayley-Hamilton theorem (for $N_{f}=2,3$).
\end{itemize}
The Schouten identity~\cite{Schouten:1938} resulting from the fact that any completely anti-symmetric tensor of rank higher than the number of dimensions must vanish, yields no extra relations here since Lorentz invariance in the nonanomalous sector implies that there can be at most four independent indices.

We first construct all operators using the building blocks of the main set
defined in section~\ref{sec:chirallag1}. Then we determine all relations on
this set of operators since it is easier to handle the single term operators
in the relations. We then rewrite the operators in terms of the $C$-even and odd operators $\mathcal{O}_i^\pm$ defined in section~\ref{sec:chirallag2} and keep
 the $C$-even part of all relations which involve only the $N_{2n}$ monomials $\mathcal{O}_i^+$.
We remove identical relations and write the remainder in matrix form
$A_{ij}\mathcal{O}_{j}^+ = 0$. The number of linearly independent equations is thus equal to the rank of $A_{ij}$, and the minimal set of monomials
at order $p^{2n}$ therefore has $N_{2n}-\textrm{rank}(A)$ elements. 

Note that also the contact terms are linearly related, so that a similar procedure must be done in order to find the minimal number of such terms.
The differences are discussed in section~\ref{sec:relations6}.

We used \textsc{FORM}~\cite{Vermaseren:2000nd} to produce the original operator basis and to produce all linear relations. The output of this was then used as input for a \textsc{C++} program calculating the rank of $A_{ij}$ by Gaussian elimination. In order to avoid numerical imprecision, we used the library \textsc{GMP}~\cite{Granlund:2016} which allows for exact arithmetic. 
Given the size of the final matrices, we implemented a method using sparse matrices allowing the Gaussian elimination to work without swapping. The general $N_f$ case was handled by choosing a number of values for $N_f$ and checking that we got the same result, we used $N_f= 17,49,73,199$ and obtained the same basis for all of them.
 
As a check, we reproduce the known numbers for NLO and NNLO in~\cite{Gasser:1983yg,Gasser:1984gg,Bijnens:1999sh}. We have also two independent implementations of the procedures as a check. A minor check is that the relations can be rewritten separately in $C$-even and $C$-odd operators.

\subsection{Partial integration or addition of total derivatives}
\label{sec:relations1}

The order $p^{2n}$ action is invariant under the addition of a total derivative to the Lagrangian. This leads to partial-integration relations. It is very difficult to be sure one has all such relations using partial integrations.
However all total derivatives can be constructed in a fashion similar to the construction of all basis operators. We construct all operators of the form $\mathcal{O}^\mu$ that are chiral and parity invariant. We can then add a relation
$\partial_\mu \mathcal{O}^\mu=0$ since we can add a term $\partial_\mu \mathcal{O}^\mu$ to the action.

One therefore has to write down all possible structures of order $p^{2n-1}$ and then take the derivative of each these. 
Generating all possible structures $\mathcal{O}^\mu$ can be implemented in the
same way as generating all possible operators $\mathcal{O}_i$ thus guaranteeing
that we have all possible partial-integration relations.

As an example at NNNLO we get the relation
\begin{align}
 0&\,=\partial^{\mu} \left\langle \nabla _{\nu }u_{\mu}\nabla ^{\nu }\nabla _{\rho}u_{\sigma} f_{+}^{\rho \sigma}\right\rangle 
 \nonumber \\
&\,= \left\langle\nabla^\mu\left( \nabla _{\nu }u_{\mu}\nabla ^{\nu }\nabla _{\rho}u_{\sigma} f_{+}^{\rho \sigma}\right)\right\rangle 
 \nonumber \\
&\, = \left\langle\nabla^{\mu}
 \nabla _{\nu }u_{\mu}\nabla ^{\nu }\nabla _{\rho}u_{\sigma} f_{+}^{\rho \sigma}\right\rangle
 +\left\langle \nabla _{\nu }u_{\mu}\nabla ^{\mu}\nabla ^{\nu }\nabla _{\rho}u_{\sigma} f_{+}^{\rho \sigma}\right\rangle
+ \left\langle\nabla _{\nu }u_{\mu}\nabla ^{\nu }\nabla _{\rho}u_{\sigma} \nabla ^{\mu}f_{+}^{ \rho \sigma}\right\rangle \, ,
\end{align}
so that one of the operators can be written in terms of the other two. 

%At NNNLO, we find in total 502 possible operators to take the derivative of, excluding all permutations of the indices, of which 28 have one index, 132 have three indices, 220 have five indices and 122 have 7 indices.

\subsection{Field redefinitions or equations of motion}
\label{sec:relations2}

Field redefintions also allow to remove terms from the Lagrangian. This is
at the level of the terms that can be removed equivalent to deriving relations
between the higher order terms using the LO EoM~\cite{Scherer:1994wi,Bijnens:1999sh}.

The LO Lagrangian in~(\ref{eq:LOLag}) has EoM
\begin{align}\label{eq:eom}
\nabla ^{\mu}u_{\mu} -\frac{i}{2}\Big( \chi _{-}-\frac{1}{N_{f}}\langle \chi _{-}\rangle  \Big) =0 \, .
\end{align}
The way we implement this is to take the set of all possible operators generated using the building blocks. We then look for all occurrences
of $\nabla^\mu u_\mu$ or covariant derivatives thereof and then use (\ref{eq:eom}) to produce a relation.

An example is the operator $\langle\chi_+u^\rho\chi_+\nabla_\rho\nabla^\mu u_\mu\rangle$ which using the EoM leads to the relation
\begin{align}
0=\langle\chi_+ u^\rho\chi_+\nabla_\rho\nabla^\mu u_\mu\rangle
-\frac{i}{2}\langle\chi_+ u^\rho\chi_+\nabla_\rho\chi_-\rangle
+\frac{i}{2N_f}\langle\chi_+ u^\rho\chi_+\rangle\langle\nabla_\rho\chi_-\rangle
\,.
\end{align}
We replace all possible occurrences of $\nabla^\mu u_\mu$ but only one at  a time if it occurs twice in an operator. Using it twice in the same term does not yield new relations. 
Since we have constructed all possible operators this method catches all possible relations from field redefinitions. 

\subsection{The Bianchi identity}
\label{sec:relations3}

The field strength tensor $\Gamma _{\mu \nu}$ is defined through the equation
\begin{align}
\label{eq:fsdef}
\left[ \nabla _{\mu},\nabla _{\nu} \right] X = \left[ \Gamma _{\mu \nu},X\right]  \, ,
\end{align}
and can be expressed in our building blocks via
\begin{align}
\label{eq:Gammamunu}
\Gamma_{\mu\nu} = \frac{1}{4}\left[ u_{\mu}, u_{\nu}\right] -\frac{i}{2}f_{+\mu \nu}\,.
\end{align}
$\Gamma_{\mu\nu}$ satisfies the Bianchi identity
\begin{align}
B_{\mu\nu\rho}\equiv\nabla _{\mu} \Gamma _{\nu \rho} +\nabla _{\nu} \Gamma _{\rho \mu}+\nabla _{\rho} \Gamma _{\mu \nu} = 0 \,.
\end{align}
The Bianchi identity is of order $p^{3}$. 
Using the definition of $\Gamma _{\mu \nu}$ this can be written
\begin{align}
\label{eq:defBBianchi}
B_{\mu\nu\rho} = \frac{1}{4}\Big( \big[ u_{\rho},f_{-\mu \nu}\big]  +\big[ u_{\mu},f_{-\nu \rho}\big]+ \big[ u_{\nu},f_{-\rho \mu}\big] \Big) -\frac{i}{2}\Big(  \nabla _{\rho} f_{+\mu \nu} + \nabla _{\mu}f_{+\nu \rho} + \nabla _{\nu} f_{+\rho \mu }\Big) \, .
\end{align}
We will use it in this form which makes it clear that in the absence of
external fields the Bianchi identity does not give extra relations.
$B_{\mu\nu\rho}$ is cyclic in $\mu\nu\rho$ and transforms as our main set
of building blocks. It is also automatically zero under a contraction of
any two indices.

Based on chiral dimension, the only structure the Bianchi identity can be traced with at NLO is $u_{\mu}$, but as each term in the Bianchi identity is of even parity (this is easily seen from table~\ref{table:transprops}) such an operator would be odd under parity. Therefore, the Bianchi identity becomes non-trivial only beyond NLO. At NNLO, it can be traced with $\nabla _{\mu} f_{+\nu \rho}$, $f_{-\mu \nu} u_{\rho}$ or $\nabla _{\mu}u_{\nu}u_{\rho}$~\cite{Bijnens:1999sh}. At NNNLO there are additional complications, since it is possible to have several traces and that it can be traced with tensors of rank three or five. However, we can treat $B_{\mu\nu\rho}$ as a separate building block and use the same methods as before to construct all possible operators involving $B_{\mu\nu\rho}$.
Each of these gives after inserting (\ref{eq:defBBianchi}) a relation between the operators $\mathcal{O}_i$.

It is sufficient to have one insertion of $B_{\mu\nu\rho}$ since a relation
involving it twice can be constructed from those only involving it once.
In addition since we have all possible partial-integration relations there is also no need to consider covariant derivatives of $B_{\mu\nu\rho}$.

\subsection{Other relations for general $N_f$}
\label{sec:relations4}

There are also other identities that can be used to obtain relations. These essentially follow from the fact that partial derivatives commute or that the commutator of partial derivatives is related to field strengths.

The first is 
\begin{align}
\label{eq:umunufmunu}
\nabla _{\mu} u_{\nu}-\nabla _{\nu} u_{\mu}+f_{-\mu \nu} = 0 \, .
\end{align}
This implies that $\nabla _{\mu}u_{\nu}$ is symmetric in its two indices when neglecting external fields. This is implemented in a very similar way as done for the EoM. We look for all occurrences of $\nabla_\mu u_\nu$ and covariant derivatives of it in our complete
list of operators and replace it with the relation (\ref{eq:umunufmunu}).

The second such equation is the defining relation of the field-strength tensor in~(\ref{eq:fsdef}) and its expression in terms of the building blocks
(\ref{eq:Gammamunu}). We implement all these relations by looking inside the list of operators for quantities that have two covariant derivatives acting on them. We then replace $\nabla_\mu\nabla_\nu X$ by
\begin{align}
0=\nabla_\mu\nabla_\nu X-\nabla_\nu\nabla_\mu X-\Gamma_{\mu\nu}X+X\Gamma_{\mu\nu}
\end{align}
and replace $\Gamma_{\mu\nu}$ by its expression (\ref{eq:Gammamunu}). This leads
to a number of relations between the original operators.
We also have cases where more than two covariant derivatives act on an object $X$. One could in principle also use the relation then on two ``inside'' covariant derivatives. This does not lead to new relations since we have included all possible operators and already have all partial-integration relations. The relation produced by ``inside'' covariant derivatives will be produced by the relation from the term where the ``outside'' derivatives have been partially integrated first.

\subsection{The Cayley-Hamilton theorem}
\label{sec:relations5}

The characteristic polynomial of any $N_{f}\times N_{f}$ matrix $A$ can be written
\begin{align}
0= p(\lambda ) = \textrm{det}\left( \lambda I - A\right) = \lambda ^{N_{f}} \, \textrm{det}\left( I-\frac{A}{\lambda}\right)  \, ,
\end{align}
where $\lambda$ is an eigenvalue of $A$ and $I$ is the unit matrix. This can be written
\begin{align}
p(\lambda) = \lambda ^{N_{f}} \, \exp \Bigg\langle \ln \left( I-\frac{A}{\lambda }\right) \Bigg\rangle \, ,
\end{align}
and when expanded in $1/\lambda$ this yields the respective polynomial coefficients in $p(\lambda)$. The Cayley-Hamilton theorem states that any $N_{f}\times N_{f}$ matrix satisfies its own characteristic equation, i.e. $p(A) = 0$ as a matrix equation. For $N_{f}=2$ and $N_{f}=3$ one finds
\begin{align}
 N_{f}=2: \; \; \; \; & A^{2}-A\langle A\rangle -\frac{1}{2}\langle A^{2}\rangle + \frac{1}{2}\langle A \rangle ^{2}  = 0  \, , 
\nonumber \\
 N_{f} = 3: \; \; \; \;  & A^{3} - A^{2}\langle A\rangle -\frac{1}{2}A\langle A^{2}\rangle +\frac{1}{2}A\langle A\rangle ^{2} - \frac{1}{3}\langle A^{3}\rangle + \frac{1}{2}\langle A\rangle \langle A^{2}\rangle -\frac{1}{6} \langle A\rangle ^{3} =0 \, .
\end{align}
For a given $N _{f}$ the matrix $A$ can be split into a sum of $N_{f}$ terms, and the Cayley-Hamilton theorem thus gives a relation between them. Letting $A = B+C$ for $N_{f}=2$ and using that both $C$ and $D$ also satisfy the Cayley-Hamilton theorem yields
\begin{align}
\label{eq:CH2}
 N_{f}=2: \; \; \; \; & \left\{ B,C\right\} -B \langle C \rangle - C \langle B \rangle -\langle BC\rangle +\langle B \rangle \langle C\rangle =0 \, . 
\end{align}
For $N_{f} = 3$ and $A = B+C+D$ the relation is
\begin{align}
\label{eq:CH3}
 N_{f}=3: \; \; \; \; & BCD +DBC +CBD+DCB +CDB +BDC  -DB \langle C\rangle
 \nonumber \\
 &-BD \langle C\rangle   -BC \langle D\rangle  - CB \langle D\rangle - DC \langle B\rangle - CD \langle B\rangle -D\langle BC \rangle 
\nonumber \\ 
& - B\langle CD\rangle -C \langle B D \rangle - \langle BCD \rangle - \langle CBD\rangle + D \langle B\rangle \langle C\rangle + B\langle C\rangle \langle D\rangle 
\nonumber \\
&+ C\langle B\rangle \langle D\rangle + \langle D \rangle \langle BC\rangle +\langle B\rangle \langle CD \rangle +\langle C \rangle \langle BD \rangle - \langle B \rangle \langle C\rangle \langle D\rangle = 0 \, .
 \end{align}
Tracing each of the above relations with any other matrix of the corresponding size thus produces a relation between the monomials.

In practice how we use these relations is that we look for all possible operators constructed using our building blocks and for each possible trace we need to look for all possibilities of having blocks $BC$. $B$ and $C$ can be a single building block but also products of them. Once we have a choice, we replace $BC$ with the relation (\ref{eq:CH2}) giving a relation between 
operators. The three-flavour case is dealt with in an analoguous manner
but looking for a possible block $BCD$ inside traces and replacing it with (\ref{eq:CH3}).

\subsection{Contact terms}
\label{sec:relations6}

As noted in section~\ref{sec:chirallag3} here we have to work in the first basis.
We classify the possible operators into three classes: Terms
\begin{enumerate}
\item with only $\chi$, $\chi ^{\dagger}$, $\tilde{\chi}$ and $\tilde{\chi}^{\dagger}$
and covariant derivatives,
\item containing both $\chi,\tilde\chi$ and field strengths.
\item with only $F_L^{\mu \nu}$ and $F_R^{\mu \nu}$ and covariant derivatives,
\end{enumerate}
For the first two classes there were sufficiently few terms that it could be handled by hand. The last class can be dealt with in the same fashion as before, but with some minor modifications. Partial integrations are immediately generalizable.
Field redefinitions are not relevant for the contact terms since the external fields are given. The Bianchi identity is valid separately for the left and right handed field strengths\footnote{These do not provide extra relations
on our main basis, they can be derived from the relations in sections~\ref{sec:relations3} and \ref{sec:relations4}.}
\begin{align}
D_\mu F_{L\nu\rho}+D_\nu F_{L\rho\mu}+D_\rho F_{L\mu\nu}&\,=0\,,
\nonumber\\
D_\mu F_{R\nu\rho}+D_\nu F_{R\rho\mu}+D_\rho F_{R\mu\nu}&\,=0\,.
\end{align}
The relations from commuting covariant derivatives split up according to the four cases in (\ref{eq:covder})
depending on the transformation property of the building block:
\begin{align}\label{eq:fsdeflr}
D_\mu D_\nu O -D_\nu D_\mu O+iF_{R\mu\nu}O-iOF_{L\mu\nu}=0,&\quad
O\longrightarrow g_ROg_L^\dagger\,,
\nonumber\\
D_\mu D_\nu O -D_\nu D_\mu O+iF_{L\mu\nu}O-iOF_{R\mu\nu}=0,&\quad
O\longrightarrow g_L O g_R^\dagger\,,
\nonumber\\
D_\mu D_\nu O -D_\nu D_\mu O+iF_{R\mu\nu}O-iOF_{R\mu\nu}=0,&\quad
O\longrightarrow g_ROg_R^\dagger\,,
\nonumber\\
D_\mu D_\nu O -D_\nu D_\mu O+iF_{L\mu\nu}O-iOF_{L\mu\nu}=0,&\quad
O\longrightarrow g_LOg_L^\dagger\,.
\end{align} 

The Cayley-Hamilton relations remain the same, however one must make sure that the (products of) building blocks that form $B,C,D$ in (\ref{eq:CH2}) or
(\ref{eq:CH3}) transform either all under $SU(N_f=2,3)_L$ or all under $SU(N_f=2,3)_R$
only. In practice this means the Cayley-Hamilton relation is only useful
for classes 2 and 3.

We wrote a separate set of programs to determine the number of independent
contact terms of class 3 and a final set to remove terms that can be removed when the minimal set of contact terms is included.

\section{The NNNLO Lagrangian}
\label{sec:nnnlolag}

In this section we present the NNNLO Lagrangian obtained from the considerations in the previous sections. The number of independent monomials
and the number of contact terms is unique. 

The original papers constructing the different Langrangians made a number of choices of which monomials to keep and which to consider superfluous. However, the actual choice of the independent set of monomials is very much a matter of choice.

We follow here a number of guiding principles to keep terms:
\begin{enumerate}
\item We always keep the maximal number of independent contact terms.
\item We remove terms that vanish when external fields vanish as much as possible.
\item We remove terms involving covariant derivatives in favour of those involving external fields.
\item We remove terms that contribute to processes with a low number of mesons as much as possible. This is done by counting occurrences of $u_\mu,\chi_-$ and $f_{-\mu\nu}$.
\item Terms involving scalar-pseudoscalar external fields are placed before those with only vector-axial-vector external fields. 
\item After that we preferentially keep terms with lower number of flavour traces. This is to make the large $N_c$ counting~\cite{tHooft:1973alw} of the monomials explicit, only leading in $N_c$ is equivalent to keeping only single
trace monomials.
\end{enumerate}
In addition in~\cite{Bijnens:1999sh} covariant derivatives were always symmetrized and $\chi_{\pm\mu}$ was used rather than $\nabla_\mu\chi_\pm$.
In the lists of terms given we can always choose instead of the monomial
listed those with symmetrized covariant derivatives (including the index  $\mu$ of $u_\mu$) and replace $\nabla_\mu\chi_\pm$ by $\chi_{\pm\mu}$.
These are always equivalent choices of monomials.

The cases given below always include the (pseudo-)Goldstone bosons but have different sets of external fields included. The cases with only external fields correspond to the columns giving the number of contact terms.

Let us first look at the case with all external fields included.
The number of independent monomials we find at each order and the number of those that are contact terms are given for general $N_f$ flavours and for $N_f=2,3$ in table~\ref{tab:resultsall}. We agree with the known results for order $p^2,p^4,p^6$ with following caveat:
in~\cite{Weber:2008} there is one more relation listed for the $p^6$ $N_f=2$ case, however if correctly rewritten in terms of $C$-even operators this relation is equivalent to an earlier one. So after correcting that we agree
there as well.
\begin{table}[t]
\begin{center}  
\renewcommand{\arraystretch}{1.2}
\begin{tabular}{|c|cc|cc|cc|}
\hline
   &\multicolumn{2}{c|}{$N_{f}$} &\multicolumn{2}{c|}{$N_{f}=3$} &\multicolumn{2}{c|}{$N_{f}=2$} \\ \hline 
      & Total & Contact & Total & Contact & Total & Contact\\ \hline
$p^2$ & 2     & 0       & 2     &  0      & 2     & 0 \\
$p^4$ & 13    & 2       & 12    &  2      & 10    & 3 \\
$p^6$ & 115   & 3       & 94    &  4      & 56    & 4 \\
$p^8$ & 1862  & 22      & 1254  &  21     & 475   & 23\\ \hline
    \end{tabular}
   \renewcommand{\arraystretch}{1}
     
    \caption{Number of monomials in the minimal basis for the case with all external fields included. Also listed is how many of them are contact terms. Our results agree with the known ones for $p^2$, $p^4$, $p^6$.}\label{tab:resultsall}
  \end{center}
\end{table}
The monomials our program produces at order $p^6$ are slightly different from those in~\cite{Bijnens:1999sh}. We find a basis that has fewer terms with only the Goldstone bosons. We have however checked that the basis given in~\cite{Bijnens:1999sh} is a minimal basis for us as well, i.e. the 115, 94, 55~\cite{Haefeli:2007ty} monomials translated into ours form a complete set.

As an example of the size of the calculations involved we give the intermediate numbers for the general $N_f$ case. The main basis consists of 9740 $C$-even monomials and we find 12444 different relations. The Gaussian elimination gave 7878 linearly independent relations, and thus a final minimal basis of 1862 monomials.

The second case that we study is when we have no scalar or pseudoscalar external fields and is given in table~\ref{tab:resultsnochi}. This case was studied at order $p^6$ in~\cite{Ruiz-Femenia:2015mia}
and in an AdS/QCD context in~\cite{Colangelo:2012ipa}. The number of operators
at orders $p^2$, $p^4$ can be easily checked against the known full Lagrangians.
The $p^6$ case for $N_f=2$ agrees with~\cite{Ruiz-Femenia:2015mia}.
\begin{table}[t]
\begin{center}  
\renewcommand{\arraystretch}{1.2}
\begin{tabular}{|c|cc|cc|cc|}
\hline
   &\multicolumn{2}{c|}{$N_{f}$} &\multicolumn{2}{c|}{$N_{f}=3$} &\multicolumn{2}{c|}{$N_{f}=2$} \\ \hline 
      & Total & Contact & Total & Contact & Total & Contact\\ \hline
$p^2$ & 1     & 0       & 1     &  0      & 1     & 0 \\
$p^4$ & 7     & 1       & 6     &  1      & 5     & 1 \\
$p^6$ & 59    & 2       & 44    &  2      & 27    & 2 \\
$p^8$ & 963   & 15      & 591   &  13     & 238   & 11\\ \hline
    \end{tabular}
   \renewcommand{\arraystretch}{1}
     
    \caption{Number of monomials in the minimal basis for the case with no scalar or pseudoscalar external fields included. Also listed is how many of them are contact terms.}\label{tab:resultsnochi}
  \end{center}
\end{table}

The third case is when we only include scalar and pseudoscalar external fields. This is given in table~\ref{tab:resultsnolr}. The results for
$p^2$ and $p^4$ can be easily determined from the known full Lagrangians.
\begin{table}[t]
\begin{center}  
\renewcommand{\arraystretch}{1.2}
\begin{tabular}{|c|cc|cc|cc|}
\hline
   &\multicolumn{2}{c|}{$N_{f}$} &\multicolumn{2}{c|}{$N_{f}=3$} &\multicolumn{2}{c|}{$N_{f}=2$} \\ \hline 
      & Total & Contact & Total & Contact & Total & Contact\\ \hline
$p^2$ & 2     & 0       & 2     &  0      & 2     & 0 \\
$p^4$ & 10    & 1       & 9     &  1      & 7     & 1 \\
$p^6$ & 62    & 1       & 48    &  2      & 27    & 2 \\
$p^8$ & 538   & 3       & 328   &  4      & 122   & 6 \\ \hline
    \end{tabular}
   \renewcommand{\arraystretch}{1}
     
    \caption{Number of monomials in the minimal basis for the case with no vector or axial-vector external fields included. Also listed is how many of them are contact terms.}\label{tab:resultsnolr}
  \end{center}
\end{table}

The final case is when we do not include any external fields. Here there are obviously no contact terms. The number of monomials at each order is given
in table~\ref{tab:resultsnoext}.
\begin{table}[t]
\begin{center}  
\renewcommand{\arraystretch}{1.2}
\begin{tabular}{|c|c|c|c|}
\hline
   &{$N_{f}$} &{$N_{f}=3$} &{$N_{f}=2$} \\ \hline 
$p^2$ & 1     & 1      & 1    \\
$p^4$ & 4     & 3      & 2    \\
$p^6$ & 19    & 11     & 5    \\
$p^8$ & 135   & 56     & 16   \\ \hline
    \end{tabular}
   \renewcommand{\arraystretch}{1}
     
    \caption{Number of monomials in the minimal basis for the case with no  external fields included. There are no contact terms in this case.}\label{tab:resultsnoext}
  \end{center}
\end{table}
The number of monomials at $p^2,p^4$ can again be easily seen from the known Lagrangians. In~\cite{Bijnens:1999sh} the monomials were chosen to follow roughly the same criteria as given above, however this was done by hand and no check whether the basis without external fields was minimal was done.
From table~2 in~\cite{Bijnens:1999sh} one sees 21 ($N_f$), 12 ($N_f=3$)  and 6
($N_f=2$)  monomials that do not vanish when all external fields are put to zero at order $p^6$. The extra relation found in~\cite{Haefeli:2007ty} reduces the $N_f=2$-number to 5 in agreement with what we obtain. One can go back to the
unpublished notes underlying~\cite{Bijnens:1999sh} or use the list of monomials and relations given explicitly in~\cite{Weber:2008} removing all external fields to derive that for general $N_f$ at $p^6$ we have 19 independent monomials, in agreement with our result.

The number of terms is in general rather large and finding relations between processes in general will be quite some work. To order $p^6$ this was done in~\cite{Bijnens:2009zd}. However we can look at the types of terms that
appear for the case with no external fields. Our ordering scheme removes terms preferentially with a lower minimal number of mesons involved.
We can thus check how many monomials require at least 4, 6 or 8 mesons in the vertices to contribute. This is done by explicitly looking at the minimal Lagrangians our programs produce. The result is given in table~\ref{tab:resultsmesons}. Note that we have a slightly lower number of monomials contributing to terms with only 4 mesons than the monomials quoted in~\cite{Bijnens:1999sh}. The number of LECs contributing to 4-meson processes is compared with a general amplitude analysis in section~\ref{sec:pipi}.
\begin{table}[t]
\begin{center}  
\renewcommand{\arraystretch}{1.2}
\begin{tabular}{|c|c|c|c|c|}
\hline
   &\#mesons&{$N_{f}$} &{$N_{f}=3$} &{$N_{f}=2$} \\ \hline 
$p^2$ & 4 & 1  & 1  & 1 \\
\hline
$p^4$ & 4 & 4  & 3  & 2 \\
\hline
$p^6$ & 4 & 4  & 3  & 2 \\
      & 6 & 15 & 8  & 3\\
\hline
$p^8$ & 4 & 6  & 5  & 3\\
      & 6 & 60 & 31 & 9\\
      & 8 & 69 & 20 & 4\\
\hline
    \end{tabular}
   \renewcommand{\arraystretch}{1}
     
    \caption{Number of monomials in the minimal basis for the case with no external fields included that produce vertices starting at the given number of mesons.}\label{tab:resultsmesons}
  \end{center}
\end{table}

\section{Meson-meson scattering}
\label{sec:pipi}

The general amplitude $M(s,t,u) = \big\langle \phi ^{c}(p_{c}) \, \phi ^{d}(p_{d}) \left| \phi ^{a}(p_{a}) \,  \phi ^{b}(p_{b}) \right. \big\rangle $ of meson-meson scattering for a general $N_{f}$ can be written as~\cite{Bijnens:2011fm,Chivukula:1992gi}
\begin{align}
 M(s,t,u) =& \Big[ \langle X^{a}X^{b}X^{c}X^{d}\rangle + \langle X^{a}X^{d}X^{c}X^{b}\rangle \Big] B(s,t,u) 
\nonumber\\
& +\Big[ \langle X^{a}X^{c}X^{d}X^{b}\rangle + \langle X^{a}X^{b}X^{d}X^{c}\rangle \Big] B(t,u,s) 
\nonumber\\
& +\Big[ \langle X^{a}X^{d}X^{b}X^{c}\rangle + \langle X^{a}X^{c}X^{b}X^{d}\rangle \Big] B(u,s,t) 
\nonumber\\
& + \delta ^{ab}\delta ^{cd} C(s,t,u)+ \delta ^{ac}\delta ^{bd} C(t,u,s)+\delta ^{ad}\delta ^{bc} C(u,s,t) \, ,
\end{align}
where $s$, $t$ and $u$ are the Mandelstam variables
\begin{align}
 s&\, = (p_{a}+p_{b})^{2}, & t &\,=(p_a+p_c)^2, & u &\,=(p_a+p_d)^2,
\end{align}
satisfying $s+t+u = 0$ in the massles case considered here.
The $X^a$ are $SU(N_f)$ generators normalized to 1.
The functions satisfy $C(s,t,u) = C(s,u,t)$ and $B(s,t,u)=B(u,t,s)$. This follows from crossing and the $SU(N_f)_V$ symmetry.

For pion-pion scattering and $N_{f}=2$ the amplitude can be written
\begin{align}
M(s,t,u) = \delta ^{ab}\delta ^{cd} A(s,t,u)+ \delta ^{ac}\delta ^{bd} A(t,u,s)+\delta ^{ad}\delta ^{bc} A(u,s,t)\, ,
\end{align}
where $A(s,t,u)=A(s,u,t)$. The traces with four generators can here be written in terms of the Kronecker delta terms.

For $N_f=3$ the Cayley-Hamilton relation (\ref{eq:CH3}) shows that the sum
of all 6 four generator traces can be rewritten in terms of the Kronecker delta terms. The part of $B(s,t,u)$ that is fully symmetric can be transferred to the $C(s,t,u)$ using that relation.

The tree level contributions from our Lagrangian become polynomials in $s,t,u$,
so we can expand the two functions as
\begin{align}
C(s,t,u) =& \alpha_0 +\alpha _{2} s
+\alpha_{41} s^2+\alpha_{42} (t-u)^2
+\alpha_{61} s^3+\alpha_{62} s(t-u)^2
\nonumber\\ &
+\alpha_{81} s^4+\alpha_{82} s^2(t-u)^2+\alpha_{83} (t-u)^4\,,
\nonumber\\ 
B(s,t,u) =& \beta_0 +\beta _{2} t
+\beta_{41} t^2+\beta_{42} (s-u)^2
+\beta_{61} t^3+\beta_{62} t(s-u)^2
\nonumber\\ &
+\beta_{81} t^4+\beta_{82} t^2(s-u)^2+\beta_{83} (s-u)^4\,.
\end{align}
Chiral invariance requires $\alpha_0=\beta_0=0$.

For $N_f=2$ we only have $A(s,t,u)$ which can be expanded as $C(s,t,u)$.
The number of free parameters at order $p^{2n}$ matches exactly that in the corresponding column of table~\ref{tab:resultsmesons}.

For general $N_f$ the number of parameters doubles w.r.t. $N_f=2$ since we now have both functions $C(s,t,u)$ and $B(s,t,u)$ present. This again matches the
numbers in table~\ref{tab:resultsmesons}.

The $N_f=3$ case is a bit more subtle, we can transfer the fully symmetric part in $s,t,u$ away from $B(s,t,u)$ so we have to determine how many of these
there are at each power taking into account $s+t+u=0$. At orders $p^4$, $p^6$, $p^8$  there is only one such combination at each order. This agrees with the numbers in the $N_f=3$ column in table~\ref{tab:resultsmesons}. If one only considers
meson-meson scattering no results follow from chiral symmetry beyond those
of $SU(N_f)_V$ in the massless limit except that $\alpha_0=\beta_0=0$.
This is not very surprising, derivative couplings in the massless limit are always soft for all legs for a four meson vertex, see~\cite{Cheung:2016drk}.

\section{Conclusions}
\label{sec:conclusions}

In this manuscript we have determined the order $p^8$ or NNNLO chiral
perturbation theory Lagrangian for the purely mesonic case of positive
intrinsic parity. We have reproduced the known results at lower orders and
discussed the cases of general number of light flavours and with two and
three light flavours. We also discussed the cases where the types of
external fields are restricted to either scalar-pseudoscalar or
vector-axial-vector as well as the case with no external fields. We separately
determined the number of contact terms.

Our main results are the number of terms given in
tables~\ref{tab:resultsall}--\ref{tab:resultsmesons}. The list of monomials is included as supplementary material but the contact terms have been listed in the appendix in table~\ref{tab:contact}.

As an example of how our results can be used we looked at meson-meson scattering and the general form of the amplitude and compared it with the freedom our Lagrangian allows.

\section*{Acknowledgements}

This work is supported in part by the Swedish Research Council grants
contract numbers 2015-04089 and 2016-05996 and by
the European Research Council (ERC) under the European Union's Horizon 2020
research and innovation programme (grant agreement No 668679). SW thanks Shangai University for supporting her stay in Lund financially.

\appendix

\section{Contact terms}\label{app:contact}
In this appendix we list the contact terms for both a general and a fixed number of flavours. These particular operators have been chosen so as to minimise the number of traces. 

For a general $N_{f}$, there are 22 contact terms, of which 15 are of class 3 and 7 belong to classes 1 and 2. For $N_{f}=3$ there are 13 class 3 monomials and 8 from classes 1 and 2, thus yielding 21 contact terms in total.  Finally, for $N_{f}=2$ the number of contact terms is 23, to which class 3 contributes 11 and the remaining 12 come from classes 1 and 2.
These are listed in
table~\ref{tab:contact}. They are ordered by class but those involving $\tilde\chi$ are put last.

\begin{table}
\begin{center}
\renewcommand{\arraystretch}{1.2}
\begin{tabular}{|c|c|c|c|}
\hline
monomial & $N_f$ & $N_f=3$ & $N_f=2$ \\
\hline
$\trace{\chi\chi^\dagger\chi\chi^\dagger}$          & 1841 & 1234 & 453 \\
$\trace{\chi\chi^\dagger}\trace{\chi\chi^\dagger}$  & 1842 & 1235 & 454 \\
$\trace{D^2\chi D^2\chi^\dagger}$                  & 1843 & 1236 & 455 \\
$i\trace{D_\mu\chi D_\nu\chi^\dagger F_R^{\mu\nu}
 +D_\mu\chi^\dagger D_\nu\chi F_L^{\mu\nu} }$       & 1844 & 1237 & 456 \\
$\trace{\chi F_{L\mu\nu} \chi^\dagger F_R^{\mu\nu}}$  & 1845 & 1238 & 457 \\
$\trace{\chi\chi^\dagger F_{R\mu\nu} F_R^{\mu\nu}+
  \chi^\dagger\chi F_{L\mu\nu} F_L^{\mu\nu}} $        & 1846 & 1239 & 458 \\
$\trace{\chi\chi^\dagger}\trace{ F_{R\mu\nu} F_R^{\mu\nu}+ F_{L\mu\nu} F_L^{\mu\nu}}$
                                                   & 1847 & 1240 & \\
$\trace{D^2 F_{L\mu\nu} D^2 F_L^{\mu\nu}}+ L\to R$ 
                                                   & 1848 & 1241 & 459 \\
$i\trace{F_{L\mu\nu}D^\rho F_L^{\mu\sigma} D_\rho F_{L~\sigma}^{~\nu}}+L\to R$
                                                   & 1849 & 1242 & 460 \\
$i\trace{F_{L\mu\nu}D^\rho F_L^{\mu\sigma} D_\sigma F_{L~\rho}^{~\nu}}+L\to R$
                                                   & 1850 & 1243 & 461\\
$\trace{F_{L\mu\nu} F_L^{\mu\nu}}\trace{F_{L\rho\sigma} F_L^{\rho\sigma}}+L\to R$
                                                   & 1851 &      & \\
$\trace{F_{L\mu\nu} F_L^{\mu\rho}}\trace{F_L^{\sigma\nu} F_{L\sigma\rho}}+L\to R$
                                                   & 1852 &      & \\
$\trace{F_{L\mu\nu} F_{L\rho\sigma}}\trace{F_L^{\mu\nu} F_L^{\rho\sigma}}+L\to R$
                                                   & 1853 & 1244 & \\
$\trace{F_{L\mu\nu} F_{L\rho\sigma}}\trace{F_L^{\mu\rho} F_L^{\nu\sigma}}+L\to R$
                                                   & 1854 & 1245 & \\
$\trace{F_{L\mu\nu} F_L^{\mu\nu}}\trace{F_{R\rho\sigma} F_R^{\rho\sigma}}$
                                                   & 1855 & 1246 & 462\\
$\trace{F_{L\mu\nu} F_L^{\mu\rho}}\trace{F_R^{\sigma\nu} F_{R\sigma\rho}}$
                                                   & 1856 & 1247 & 463\\
$\trace{F_{L\mu\nu} F_{L\rho\sigma}}\trace{F_R^{\mu\nu} F_{R}^{\rho\sigma}}$
                                                   & 1857 & 1248 & 464\\
$\trace{F_{L\mu\nu} F_{L\rho\sigma}}\trace{F_R^{\mu\rho} F_{R}^{\nu\sigma}}$
                                                   & 1858 & 1249 & 465\\
$\trace{F_{L\mu\nu}F_{L}^{\mu\nu}F_{L\rho\sigma}F_{L}^{\rho\sigma}}+L\to R$
                                                   & 1859 & 1250 & 466\\
$\trace{F_{L\mu\nu}F_{L}^{\mu\rho}F_{L}^{\nu\sigma}F_{L\rho\sigma}}+L\to R$
                                                   & 1860 & 1251 & 467\\
$\trace{F_{L\mu\nu}F_{L}^{\mu\rho}F_{L\rho\sigma}F_{L}^{\nu\sigma}}+L\to R$
                                                   & 1861 & 1252 & 468\\
$\trace{F_{L\mu\nu}F_{L\rho\sigma}F_{L}^{\mu\nu}F_{L}^{\rho\sigma}}+L\to R$
                                                   & 1862 & 1253 & 469\\
\hline
$\trace{D_\mu \chi D^\mu \tilde\chi^\dagger+D_\mu\chi^\dagger D^\mu\tilde\chi}$
                                                   &      & 1254 & \\
\hline
$\trace{\tilde\chi\chi^\dagger\chi\chi^\dagger+\tilde\chi^\dagger\chi\chi^\dagger\chi}$
                                                    &      &     & 470\\
$\trace{\tilde\chi\chi^\dagger\tilde\chi\chi^\dagger+\tilde\chi^\dagger\chi\tilde\chi^\dagger\chi}$
                                                    &      &     & 471\\
$\trace{D^2\chi D^2\tilde\chi^\dagger+D^2\chi^\dagger D^2\tilde\chi}$
                                                    &      &     & 472\\
\parbox{6.5cm}{$i\trace{\left(D_\mu\chi D_\nu\tilde\chi^\dagger +D_\mu\tilde\chi D_\nu\chi^\dagger\right)F_R^{\mu\nu}}$ \\
$~~~~+i\trace{
\left(D_\mu\chi^\dagger D_\nu\tilde\chi +D_\mu\tilde\chi^\dagger D_\nu\chi\right)F_L^{\mu\nu}}$}
 & & & 473\\
$\trace{\chi F_{L\mu\nu}\tilde\chi^\dagger F_R^{\mu\nu}+\tilde\chi F_{L\mu\nu}\chi^\dagger F_R^{\mu\nu}}$ & & & 474\\
\parbox{5.5cm}{
$\trace{\left(\chi\tilde\chi^\dagger+\tilde\chi\chi^\dagger\right)F_{R\mu\nu}F_R^{\mu\nu}}$\\
$~~~~+\trace{\left(\chi^\dagger\tilde\chi+\tilde\chi^\dagger\chi\right)F_{L\mu\nu}F_L^{\mu\nu}}$} & & & 475\\
\hline
\end{tabular}
\end{center}
\caption{The list of contact terms and the number in the minimal basis
for a general number of flavours $N_f$ and for $N_f=2,3$.}
\label{tab:contact}
\end{table}

\bibliography{p8lagrangian}
\bibliographystyle{JHEPmy}

\end{document}